\documentclass[12pt]{article}
\usepackage[breaklinks=true]{hyperref}
\usepackage{graphicx}
\usepackage{epsfig}
\usepackage{dcolumn}
\usepackage{subfigure}
\usepackage{amsmath,amsthm, amssymb}
\usepackage{slashed}

\usepackage{comment}

\newcommand{\bea}{\begin{eqnarray}}
\newcommand{\eea}{\end{eqnarray}}
\newcommand{\beq}{\begin{equation}}
\newcommand{\eeq}{\end{equation}}
\newcommand{\nn}{\nonumber}

\newcommand{\BB}[1]{{\mathbb{#1}}}

\newcommand{\half}{{\frac{1}{2}}}

\newcommand{\braket}[2]{\langle #1\vert\,#2\rangle}

\newcommand{\e}{\ \mbox{e}}
\newcommand{\la}{\lambda}
\newcommand{\al}{\alpha}
\newcommand{\hal}{{\hat{\alpha}}}
\newcommand{\bCH}{\bar{\C{H}}}

\def\be{\begin{equation}}
\def\ee{\end{equation}}
\def\beq{\begin{eqnarray}}
\def\eeq{\end{eqnarray}}

\newcommand{\C}[1]{\mathcal{#1}}
\newcommand{\sect}[1]{\setcounter{equation}{0}\section{#1}}

\begin{document}

\title{\Large{{\bf
Higher Order Analogues of Tracy-Widom Distributions via the Lax Method
}}}

\vspace{1.5cm}

\author{~\\{\sc Gernot Akemann} and {\sc Max R. Atkin}
\\~\\
Fakult\"{a}t f\"{u}r Physik, Universit\"{a}t Bielefeld,\\
%Department of Physics, Bielefeld University,\\
Postfach 100131,
D-33501 Bielefeld, Germany\\~\\
}
\date{}
\maketitle
\vfill
\begin{abstract}
We study the distribution of the largest eigenvalue in Hermitian one-matrix
models when the spectral density acquires an extra
number of $k-1$ zeros at the edge.
The distributions are directly expressed through the norms of orthogonal
polynomials on a semi-infinite interval, as an alternative to
using Fredholm determinants.
They satisfy non-linear recurrence
relations which we show form a Lax pair, making contact to the
string literature in the early 1990's. The technique of pseudo-differential
operators allows us to give compact expressions for the logarithm
of the gap probability
in terms of the Painlev\'e XXXIV hierarchy.
These are the higher order analogues
of the Tracy-Widom distribution which has $k=1$. Using known B\"acklund
transformations we show how to simplify earlier
equivalent results that are derived from Fredholm determinant theory,
valid for even $k$ in terms of the Painlev\'e~II hierarchy.
\end{abstract}
%PACS

\vfill

\thispagestyle{empty}
\newpage

\renewcommand{\thefootnote}{\arabic{footnote}}
\setcounter{footnote}{0}

%%%%%%%%%%%%%%%%%%%%%%%%%%%%%%%%%%%%%%%%%%%%%%%%%%%%%%%%%%%%%%%%%%%%%%%%%%%
\sect{Introduction}\label{intro}
Random matrices appear in many branches of mathematics, physics and other
sciences. Perhaps the best example to illustrate such a breadth of applications
is a single formula known as the Tracy-Widom (TW) distribution, and we refer to
\cite{TWreview} for an incomplete list. It was first realised
as the scaled distribution of the largest eigenvalue of a large complex
Hermitian random matrix with Gaussian entries \cite{TW}
that can be expressed as the derivative of the
Fredholm determinant of the Airy-kernel
\cite{PF93}. Apart from its appearance in many other areas this distribution
is remarkably robust in the sense that it holds for a much larger class
of non-Gaussian distributions, of either invariant \cite{Deift} or Wigner type
\cite{Soshnikov}; this property is known as universality.

Another striking feature is the relation between the TW
distribution and integrable hierarchies
of non-linear differential equations, notably the Painlev\'e hierarchies.
A relation between matrix models and integrable hierarchies was first observed in the string literature of the early 1990's
where matrix models were applied to two-dimensional quantum gravity, and we
refer to \cite{PDFGZ} for an extensive review.
There the possibility of having ``multicritical points'' was introduced in
order to couple matter to gravity. This happens when $k-1$ additional zeros
collide with the spectral density - which is the semi-circle in the Gaussian
case ($k=1$).
Such scenarios can be realised by fine-tuning non-Gaussian potentials to ``critical potentials'',
see e.g. \cite{Moore,DJM} for a classification and references.
In order to have the largest possible set of
critical exponents available this also included so-called formal matrix models,
where the confining potential is not bounded from below. This allows $k$ to assume any integer value.
When considering multicritical models it is usual to study the perturbations of such models away from
criticality by suitably scaling some of the parameters in the potential as it approaches its critical form.
It should be noted that sometimes in the mathematics literature the term multicritical model is reserved for the case when all possible perturbations are included, however in this article we will use the term multicritical in the sense introduced above; this coincides with the physics literature.

Whereas the limiting kernel of orthogonal polynomials with multicritical
potentials has been considered before in
mathematics, e.g. in \cite{TM,Adler1}, and
in the physics literature, see \cite{BM}
for a comprehensive list of references,
it was not until recently that the question about the existence
of higher order analogues of TW distributions at such multicritical points
was first answered \cite{TC1}. This also includes a first numerical
investigation in \cite{TC2}.
Building on the Riemann-Hilbert analysis
of the
limiting kernel for $k=3$ \cite{TM}, in \cite{TC1} Fredholm determinant theory
was used to describe the higher order analogues of TW for any odd $k$
\footnote{Note that our convention of counting zeros differs from \cite{TC1}.}.
They were found to be characterised by a system of equations including the odd
members of the Painlev\'e II hierarchy.

Several alternative derivations of the TW distributions exist, in particular
a recent heuristic but very transparent one \cite{SN}, that avoids the Fredholm
determinant formalism. Starting from a Gaussian potential they directly
compute the gap probability that the interval $(y,\infty)$ is empty of
eigenvalues. It is given in terms of the norms of orthogonal polynomials
on the complement  $(-\infty,y]$ which satisfy a set of non-linear differential
equations. After carefully performing the asymptotic analysis of the
recursion coefficients in the three-step recurrence relation
TW at $k=1$ follows.

Our goal is to generalise the approach \cite{SN}
to the higher order multicritical
cases. However, it rapidly becomes very cumbersome, already when extending
their derivation of TW to a generic quartic potential (which only shows its
known universality). The main idea in this paper is to cast the recurrence
relations into a different
set of flow or string equations, that allow to make close
contact to integrable hierarchies via the Lax pair. Once these
commutation relations become multiplicative we can apply the Lax method to
solve them using  pseudo-differential operators, following \cite{PDFGZ,Das}.
Ultimately this leads to
a hierarchy of differential equations given by the Painlev\'e XXXIV
hierarchy for any integer $k$,  in terms of which
the logarithm of the gap probability at multicriticality can be
expressed.
Here we exploit that the limiting case $y\to\infty$ is well understood in
terms of these integrable hierarchies, see e.g. \cite{PDFGZ}.
Furthermore, we manage to prove the equivalence to the subset of odd-$k$
hierarchies that were derived rigorously in \cite{TC1}. The link is provided
by a B\"acklund transformation
that was explicitly constructed in \cite{Clarkson}.
Applied to our result it leads to all members of Painlev\'e II including
\cite{TC1}.
A shift property of the so-called Lenard differential
operators (also known as Gelfand-Dikii polynomials) allows us to simply
their result.

%While completing this work a paper has appeared \cite{Adler2} that also exploits the relation between the integrable and string literature for multicritical points, in order to arrive at alternative representations for the higher order analogues of TW {\bf comment in more detail on their results}.

The remainder of this paper is organised as follows.
In section \ref{OP} we review the orthogonal polynomial formalism adopted
to \cite{SN} and derive our flow and string equation for finite matrix size.
The main result is derived in section \ref{Lax} where the double scaling
limit is taken on the string equation, leading to Painlev\'e hierarchies
for the gap probability. Explicit examples including TW are given and the
equivalence to \cite{TC1} is shown, before concluding in section \ref{conc}.
Various technical details are delegated to the appendices.

%%%%%%%%%%%%%%%%%%%%%%%%%%%%%%%%%%%%%%%%%%%%%%%%%%%%%%%%%%%%%%%%%%%%%%%%%%%
\sect{The Orthogonal Polynomial Formalism}\label{OP}

Consider the following truncated partition function
\beq
Z_N(y; \alpha, \{g_l\}) &\equiv&  \frac{1}{N!}  \int^{y}_{-\infty}
\prod^{N}_{i=1} d\lambda_i \e^{-N \alpha V(\lambda_i)}
\prod^{N}_{k>j}(\la_k-\la_j)^2\ ,
\label{Zydef}\\
V(\lambda)&\equiv&\sum_{l=1}^\infty \frac{1}{l}g_l\la^l
\label{Vdef}\ .
%\Delta(\lambda)^2
\eeq
Here $V(\lambda)$ is a formal power series and $\al>0$ is a real parameter.
Both are independent of $y$.
When sending the upper integration range $y\to\infty$ it corresponds to the
standard Hermitian one-matrix model $Z_N(\infty; \alpha, \{g_l\})$,
given in terms of eigenvalues $\la_i$ of a random matrix $M$.
In general we want to compute the probability
$\BB{P}_N(\lambda_{\mathrm{max}} < y ;\alpha,\{g_l\})$ that the
largest eigenvalue of the random matrix $M$, is less than $y$.
It has the following expression,
\beq
\label{gapProbZ}
\BB{P}_N(\lambda_{\mathrm{max}} < y ;\alpha,\{g_l\})
= \frac{Z_N(y; \alpha, \{g_l\}) }{Z_N(\infty; \alpha, \{g_l\}) },
\eeq
which is why our definition eq. (\ref{Zydef}) is convenient. One possibility
to determine this probability is to introduce orthogonal polynomials
for the partition function
$Z_N(\infty; \alpha, \{g_l\})$ and to study their asymptotic
behaviour in an appropriate scaling limit. However, we will follow
\cite{SN} here by introducing polynomials
for the truncated partition $Z_N(y; \alpha, \{g_l\})$. Formally this could
also be formulated in the one-matrix model by adding a hard wall at $y$ to the
potential.
Introduce a set of polynomials $\{\pi_n(\lambda): n \in \BB{N}\}$ such
that $\pi_n(\lambda)$ is of order $n$ and they are orthonormal with respect
 to the inner product defined by,
\beq
\braket{\pi_n}{\pi_m} \equiv \int^y_{-\infty} d\lambda
\e^{-N \alpha V(\lambda)} \pi_n(\lambda)\pi_m(\lambda) = \delta_{nm} \ .
\label{OPdef}
\eeq
Note that the coefficients in the expression for each $\pi_n$ depend
on $y$, $\al$ and the potential. Furthermore, we define $h_n=h_n(y)>0$ by the
leading coefficients
\beq
\pi_n(\lambda) = \frac{1}{\sqrt{h_n}}\ \lambda^n  + O(\lambda^{n-1})\ .
\label{hndef}
\eeq
They are positive being the squared norms of the polynomials in monic
normalisation. It can be shown using the orthonormality that \cite{Mehta}
\beq
Z_N(y; \alpha, \{g_l\}) = \prod^{N-1}_{i=0} h_i = h_0^N \prod^{N-1}_{i=1}
r_i^{N-i}\ .
\label{Znorm}
\eeq
Here we have defined the ratios
\beq
\label{rndef}
r_n = \frac{h_n}{h_{n-1}} \quad \mathrm{for} \quad  n\geq 1\ ,
\eeq
that will determine the distribution of the largest eigenvalue. It is useful
to rewrite\newline
$Z_N(y; \alpha, \{g_l\})$ as,
\beq
\label{logZ}
\log[ Z_N(y; \alpha, \{g_l\})] = N \log [h_0] + N \sum^{N-1}_{i=1}
\left (1-\frac{i}{N} \right)
\log[r_i].
\eeq
In the case of a Gaussian potential $g_l = 2\delta_{2,l}$, and $y\to\infty$
everything is explicitly known for finite-$N$ because the $\pi_n$ are
proportional to Hermite polynomials, with $h_n=n!\sqrt{\pi}/2^n$ and the
ratio assuming the simple form $r_n=n/2$ (we have set $N\al=1$ for simplicity).

In order to determine the $r_i$ in a general setting we first introduce a set
of multiplication and differentiation operators on the ring of polynomials
$\pi_n$ which form a complete set of functions. We begin with the
multiplication operator $B$:
\beq
B_{nm} \pi_m(\lambda)\equiv \lambda \pi_n(\lambda) = \sqrt{r_{n+1}}
\pi_{n+1}(\lambda) + s_n \pi_n(\lambda) +\sqrt{r_{n}} \pi_{n-1}(\lambda) \ ,
\label{Bdef}
\eeq
where we have used summation conventions for the matrix multiplication on the
left hand side. In terms of the inner product this reads
\beq
B_{nm} = \braket{\pi_n }{\lambda \pi_m}= \sqrt{r_{n+1}}\,\delta_{n+1,m}
+ s_n\delta_{n,m} + \sqrt{r_{n}}\, \delta_{n-1,m}\ .
\label{Bnmdef}
\eeq
The fact that $B$ is tridiagonal follows from the three step recurrence
relation that arbitrary orthogonal polynomials with weights on the real
line satisfy; this includes the case of the truncated integral in
\eqref{Zydef}. Besides $r_n$ we encounter here a second recurrence
coefficient, $s_n=\braket{\pi_n }{\lambda \pi_n}$, that must be
determined in
principle. Obviously $B$ is a symmetric matrix as
$B_{nm} = \braket{\pi_n }{\lambda \pi_m}=\braket{\pi_m }{\lambda \pi_n}$,
or in operator language $B=B^T$.

Differentiation with respect to the argument of the polynomials is denoted
by $A$:
\beq
A_{nm} \pi_m(\la) \equiv \partial_\lambda \pi_n(\la) \ ,
\label{Adef}
\eeq
or in terms of the inner product
$A_{nm} = \braket{\pi_m }{\partial_\lambda \pi_n}$.
Because the coefficients $h_n$ are $\la$-independent the polynomial
on the right hand side is at most of degree $n-1$, and thus
\beq
\label{Alimits}
A_{n,n+k} = 0 \quad \mathrm{for} \quad k > -1 \ ,
\eeq
or in other words $A$ is strictly lower triangular. In particular for the
uppermost non-zero diagonal we have
\beq
A_{n,n-1}=\int^y_{-\infty} d\lambda \e^{-N\alpha V(\lambda)} \pi_{n-1}
\partial_\lambda\left(\frac{1}{\sqrt{h_n}}\ \lambda^n  + O(\lambda^{n-1})
\right) =n/\sqrt{r_n}\ .
\label{Aid}
\eeq
For the Gaussian potential other explicit expressions can be derived.
It is easy to see that
$A$ and $B$ form canonical commutation relations
\beq
\label{BAcommute}
[B,A] = 1 \ ,
\eeq
due to $(BA-AB)_{nm}\pi_m=\partial_\lambda (\la\pi_n)-\la\partial_\lambda
\pi_n=\pi_m\delta_{n,m}$, upon using the definitions.

Finally we also introduce differentiation with respect to the truncation
$y$ denoted by $C$:
\beq
C_{nm} \pi_m(\la) \equiv \partial_y \pi_n(\la) \ ,
\label{Cdef}
\eeq
or equivalently $C_{nm} = \braket{\pi_m }{\partial_y \pi_n}$.
Now the leading order coefficient containing $h_n$ in eq. (\ref{hndef})
is $y$-dependent,
and so the polynomial on the right hand side is at most of degree $n$,
\beq
\label{Climits}
C_{n,n+k} = 0 \quad \mathrm{for} \quad k > 0 \ .
\eeq
The matrix $C$ is lower triangular including the diagonal, which is given by
\beq
C_{nn}=\int^y_{-\infty} d\lambda \e^{-N\alpha V(\lambda)} \pi_{n}
\partial_y\left(\frac{1}{\sqrt{h_n}}\ \lambda^n  + O(\lambda^{n-1})
\right) =-\frac12\partial_y\log[h_n]\ .
\label{Cnn}
\eeq
For further
explicit expressions for the matrix elements see eq. \eqref{id1} below.
Matrix $C$ satisfies the
following commutation relations with the multiplication operator $B$
\beq
\label{BCcommute}
[B,C] = -\partial_y B\ ,
\eeq
where we have been careful to remember the $y$-dependence
of the recursion coefficients,
$(BC-CB)_{nm}\pi_m=B_{nl}\partial_y\pi_l-\partial_y (\la\pi_n)
=B_{nl}\partial_y\pi_l-\partial_y (B_{nl}\pi_l)$.
We are now prepared to introduce two composed operators that will play a
crucial role in the following, by establishing a link to integrable systems.

%%%%%%%%%%%%%%%%%%%%%%%%%%%%%%%%%%%%%%%%%%%%%%%%%
\subsection{The Flow Equation}

We begin with the matrix $P$,
\bea
\label{Pdef}
P_{nm} \equiv A_{nm} + C_{nm} -\frac{\alpha N}{2} V'(B)_{nm}
= \braket{\pi_m }{\left(\partial_\lambda + \partial_y
-\frac{\alpha N}{2} V'(\lambda) \right) \pi_n}.
\eea
It is a simple matter of integration by parts to see that $P$ is
anti-symmetric, $P+P^T=0$:
\bea
P_{nm} &=&
\int^y_{-\infty} d\lambda \e^{-N\alpha V(\lambda)} \pi_m
\left(\partial_\lambda + \partial_y
- \frac{\alpha N}{2} V'(\lambda)\right) \pi_n \nn\\
&=& \e^{-N\alpha V(y)} \pi_m(y) \pi_n(y) - \int^y_{-\infty} d\lambda
\e^{-N\alpha V(y)} \left(-{N\alpha}  V'(\lambda)  \pi_m \pi_n+
\pi_n\partial_\lambda \pi_m  \right)\nn\\
&&+ \int^y_{-\infty} d\lambda\e^{-N\alpha V(\lambda)}
\left(  \partial_y \left( \pi_n \pi_m\right) - \pi_n \partial_y \pi_m
-\frac{\alpha N}{2} V'(\lambda)\pi_m\pi_n
\right)\nn\\
&=& - \int^y_{-\infty} d\lambda \e^{-N\alpha V(y)} \pi_n\left(
\partial_\lambda + \partial_y
-\frac{N\alpha}{2}  V'(\lambda) \right)\pi_m=-P_{mn} \ .
\eea
Here we have used the fact that,
\beq
0&=&\partial_y\delta_{n,m}=
\partial_y \braket{\pi_n}{\pi_m} = \e^{-N\alpha V(y)} \pi_n(y) \pi_m(y)
+ \int^y_{-\infty} d\lambda \e^{-N\alpha V(\lambda)}\partial_y
\left(  \pi_n \pi_m\right) \nn\\
\Leftrightarrow 0&=&\e^{-N\alpha V(y)} \pi_n(y) \pi_m(y)+C_{mn}+C_{nm}\
\label{id1}
\eeq
which we have also expressed in terms of the matrix elements of $C$.
Secondly, using \eqref{Alimits} and \eqref{Climits}
we have that $P_{n,n+k} = -\frac{N\al}{2} V'(B)_{n,n+k}$ for all $k > 0$,
and hence using the anti-symmetry of $P$ we find that
\beq
P = -\frac{N\al}{2} (V'(B)_+ - V'(B)_-)\ ,
\label{Pfinal}
\eeq
where $+$ and $-$ denote the upper and lower triangular parts of a matrix
respectively.
Obviously its diagonal part vanishes, and its antisymmetry (given the
symmetry of $B$) is now manifest. Because we have expressed $P$ only in
terms of powers of $B$, which is tridiagonal,
we see $P$ has only a finite number of non-zero off diagonals. For the
Gaussian potential
with $g_l=2\delta_{2,l}$ (that is $V(\la)=\la^2$)
for example it reads
\beq
P_{nm}|_{Gauss}= -{N\al} (\sqrt{r_{n+1}}\,\delta_{n+1,m}-
\sqrt{r_n}\,\delta_{n-1,m})\ .
\label{PGauss}
\eeq
Finally, given the commutation relations \eqref{BAcommute} and
\eqref{BCcommute} we arrive at
\beq
\label{floweqn}
[P,y-B] = \partial_y (y-B) =1-\partial_y B .
\eeq
This is our first main result of this section, and we will refer to it as
the flow
equation. Note that this equation implies that the matrices $y-B$ and $P$ form
a Lax pair 
\footnote{After changing to light cone coordinates $x_{\pm}=(y\pm \lambda)/2$ and defining $\hat{B}=y-B$ the eqs. (\ref{Bdef}), (\ref{Pdef}) and (\ref{floweqn}) can be written as standard Lax eqs. $\hat{B}\psi_n=2x_-\psi_n$, $P\psi_n=\partial_+\psi_n$ and $\partial_+\hat{B}=[P,\hat{B}]$ where 
$\psi_n=\pi_n\exp[-N\alpha V/2]$.}.
%{We further emphasise this fact by noting that by letting $B$ and $P$ act on %a vector $\Psi$ we may write the flow equation as the compatibility condition %of the two equations $B \Psi = \lambda \Psi$ and $P \Psi = \partial_y \Psi$.}, 
where we have added a trivial zero, $0=[P,y]$,
to bring it to that form.
The flow equation \eqref{floweqn} provides explicit relations among the
recurrence coefficients $r_n$ and $s_n$, including derivatives with respect
to $y$. To give an example we again consider the Gaussian case.
Inserting \eqref{Bnmdef} and $P$ in the form of
\eqref{PGauss} into \eqref{floweqn} we obtain
five equations. Those for $m=n\pm2$ are identically satisfied, the
equations for $m=n\pm1$ are the same, and thus we are left with the
following two equations for indices $m = n-1$
and $m = n$:
\bea
s_{n} - s_{n-1} &=& - \frac{1}{2\alpha N} \partial_y \log [r_{n}]\ ,
\label{snG}\\
r_{n+1} - r_{n} &=& \frac{1}{2\alpha N}(1-\partial_y s_n)\ .
\label{rnG}
\eea
The first of this set of equations already appeared in \cite{SN}, however the
second is new. The equivalence of our flow equations to \cite{SN} is
shown in Appendix A. The advantage of our approach
is that we may obtain such recursion relations for any potential.
Furthermore, we can see that our recursion relations
%correspond to
enjoy an explicit link to
integrable systems defined by the Lax pair $y-B$ and $P$. Indeed we
can see this connection in this particular example
as it is trivial to eliminate $s_n$, thereby obtaining,
\beq
r_{n+1}+r_{n-1}-2r_{n} = \frac{1}{4 \alpha^2 N^2} \partial^2_y \log[ r_n]\ .
\label{Toda}
\eeq
By substituting $r_n = \exp[ \phi_n]$ we see that this is equivalent
to the difference of the Toda lattice equation at two neighbouring lattice
sites.

%%%%%%%%%%%%%%%%%%%%%%%%%%%%%%%%%%%%%%%%%%%%%%%%%
\subsection{The String Equation}

The second operator we introduce is the matrix $H$. It will lead to a
so-called string
equation that is algebraic in $r_n$ and $s_n$. It is defined as,
\beq
H_{nm}&\equiv& (A(B-y))_{nm} - \frac{N \alpha}{2} (V'(B)(B-y))_{nm}
+\frac12\delta_{nm}\nn\\
&=&\int^y_{-\infty} d\lambda \e^{-N\alpha V(\lambda)} \pi_m \left(
(\la-y)\Big(\partial_\lambda - \frac{\alpha N}{2} V'(\lambda)\Big)
+\frac12\right) \pi_n \ .
\label{Hdef}
\eeq
It is easy to see that $H$ is antisymmetric, $H+H^T=0$,
by applying the following identity in terms of a total derivative:
\beq
0&=&\int^y_{-\infty} d\lambda \partial_\lambda \left[(\lambda - y)
\e^{-N\alpha V(\lambda)} \pi_n(\lambda) \pi_m(\lambda) \right]\nn\\
&=&\int^y_{-\infty} d\lambda \e^{-N\alpha V(\lambda)} \left[
 \pi_n\pi_m+(\lambda - y)(-N\al V'(\la) \pi_n\pi_m+ (\partial_\la\pi_n)\pi_m
+\pi_n(\partial_\la\pi_m))\right]\nn\\
&=& H_{nm}+H_{mn}\ .
\eeq
Because $A$ is strictly lower triangular and $B$ tridiagonal we have that
its product does not contribute to the $+$ part of $H$, hence
$H_{n,n+k}= -\frac{N\al}{2}(V'(B)(B-y))_{n,n+k}$ for all $k > 0$. Due to the
antisymmetry of $H$ we thus arrive at
\beq
H=-\frac{N\al}{2}\Big( (V'(B)(B-y))_+- (V'(B)(B-y))_-\Big)\ .
\label{Hfinal}
\eeq
This shows that also $H$ only has a finite number of non-zero off-diagonals,
being a multiple of $B$ in its $+$ and $-$ part. For the Gaussian potential
it reads
\beq
H_{nm}|_{Gauss}&=&-{N\al}(\sqrt{r_{n+2}r_{n+1}}\,\delta_{n+2,m}
+ \sqrt{r_{n+1}}(s_{n+1}+s_n-y)\delta_{n+1,m})\nn\\
&&+{N\al}(\sqrt{r_{n}}(s_{n}+s_{n-1}-y)\delta_{n-1,m}
+\sqrt{r_{n}r_{n-1}}\,\delta_{n-2,m})\ .
\label{HGauss}
\eeq
Furthermore, from the definition \eqref{Hdef} and the
commutation relations \eqref{BAcommute} and \eqref{BCcommute} we obtain
\beq
[B-y,H]=B-y\ ,
\label{stringeqn}
\eeq
which is the second result of this section (where we have added a trivial
zero, $0=[-y,H]$). We shall refer to it as the string equation and it
provides relations among the recurrence coefficients, for any potential $V$,
which are algebraic and no longer contain derivatives.
Let us give again the Gaussian example. Plugging eq. \eqref{Bnmdef} and
\eqref{HGauss} into the string equation we obtain five equations,
two of which for
$m=n\pm2$ are identically satisfied. The ones for $m=n\pm1$ are identical and
we are left with the two equations for $m=n+1$ and $n=m$ which read,
\beq
s_{n+1}^2-s_n^2+ r_{n+2}-r_n - y (s_{n+1}-s_n)&=&\frac{1}{N\al}\ ,
\label{string1}\\
r_{n+1}(s_{n+1}+s_n-y) -r_{n}(s_{n}+s_{n-1}-y)
&=&\frac{1}{2N\al}(s_n-y)\ .
\label{string2}
\eeq
Finally we note that the two operators $P$ and $H$ are related. Because
multiplication of matrices and taking their upper (+) or lower (-) part in
general  do not commute we derive the following relation in Appendix \ref{tri}:
\beq
H&=& (B-y)P -(BP)_{\rm d} -(B_+-B_-)\frac{N\al}{2}V'(B)_{\rm d}\nn\\
&=& \frac12\{B,P\}-yP-\frac{N\al}{4}\{B_+-B_-,V'(B)_{\rm d}\}\ ,
\label{Hexp}
\eeq
where the subscript d denotes the diagonal part of a matrix. In the second
line we have made the anti-symmetry of $H$ manifest, by decomposing it into
its (vanishing) symmetric part and its anti-symmetric part.
This relation will be important when taking the double scaling limit in the
next section.

%%%%%%%%%%%%%%%%%%%%%%%%%%%%%%%%%%%%%%%%%%%%%%%%%%%%%%%%%%%%
\sect{Double Scaling Limit and the Lax Method}\label{Lax}

In this section we will take the large-$N$ limit and determine the asymptotic
recurrence coefficients. This will give the distribution of the largest
eigenvalue by differentiating \eqref{gapProbZ}. In the case of a Gaussian potential and
$y\to\infty$ it is well known that the limiting spectral density
$\rho(\la)$ is given by the Wigner semi-circle that vanishes as a
square root at its end points. By taking the large-$N$ limit while
simultaneously zooming into the vicinity of the right end point and
letting $y$ approach the same end point - hence the name double
scaling limit (d.s.l) - the distribution of the largest eigenvalue
can be found. In \cite{SN} this was done using in an approach using the
truncated partition function \eqref{Zydef} which we generalise here.

For non-Gaussian potentials it is well known \cite{Moore,DJM} how to
construct multicritical potentials $V_{k,k'}$, such that the spectral
density behaves as $\rho(\la)\sim(2a_+-\la)^{k-1/2}(\la-2a_-)^{k'-1/2}$.
Here $(2a_-,2a_+)$ is the support of $\rho$ which we will assume to be of
single interval or one-arc type in all the following, and the integers
$k,k'=1,2,\ldots$ label the degree of multicriticality, following ref.
\cite{DJM}. So for the Gaussian potential we have $(k,k')=(1,1)$ with
$a_+=a_-=1$ \footnote{Note that in \cite{TC1} the Gaussian case is
labelled by $l=k-1=0$, and that only critical points with our $k$ taking
odd values are considered.}. For these critical points a non-trivial d.s.l
can be taken if we make the following scaling ansatz \cite{Douglas},
\bea
\xi &=& \xi_c - a^k x \label{yinfscaleansatz} \\
N &=& a^{-(2k+1)/2} \ \ ,\nn
\eea
where $a$ is the scaling variable given in fractional powers of $N$ and
$n/N\to\xi$ becomes a continuous variable \footnote{Note that we always
have $\xi_c = 1$. This will be justified later.}. In addition we make the
following general ansatz for our recurrence coefficients $r_n$ and $s_n$:
\bea
%\label{rsscaling1}
r(\xi,y) &=& \ r_c(1 + a \rho_1(x,s) + O(a^{3/2}) )\ , \nn\\
s(\xi,y) &=& \sqrt{r_c}(s_c + a \sigma_1(x,s) + O(a^{3/2})) \ .\label{rsscaling2}
\eea
It will also be useful to introduce the functions,
\beq
%u_i\equiv\rho_i+\sigma_i\ ,\ \ v_i\equiv\rho_i-\sigma_i\ .
u_1\equiv\rho_1+\sigma_1\ ,\ \ v_1\equiv\rho_1-\sigma_1\ .
\label{uvdef}
\eeq
In \cite{Deift} it was shown that for a one-arc support and analytic
potentials the $r_n$ and $s_n$ indeed become single valued functions.
We assume in the following that this remains true for a one-arc support with finite $y$. The
d.s.l in the case $y = \infty$ is fully reviewed in Appendix
\ref{standarddsl}. However, to quickly recap here; by substituting the
above scaling ansatz into the $y = \infty$ string equations found in
Appendix \ref{standarddsl}, or alternatively using pseudo-differential
operators to solve the same string equation, one finds the result assuming $k' < k$,
\bea
\label{yinfscaledstr1}
\C{L}_{k}[u_1(x,\infty)] &=& x/2, \\
\label{yinfscaledstr2}
\C{L}_{k'}[v_1(x,\infty)] &=& 0.
\eea
The $\C{L}_{k}$ are the Lenard differential operators introduced in the next
subsection, see \eqref{Lenardrelation}.
In the remainder of this section we argue that the scaling ansatz given in
%\eqref{rsscaling1} and
\eqref{rsscaling2} together with \eqref{yinfscaleansatz}, augmented to,
\bea
\xi &=& \xi_c - a^k x \nn \\
y &=& y_c + a c_1 s \label{scaleansatz} \\
N &=& a^{-(2k+1)/2} \ \ ,\nn
\eea
where $\xi_c=1$, $c_1$ is an arbitrary constant to be fixed later,
and $s$ is the microscopic variable that describes the distribution of the
largest eigenvalue around the right endpoint of the support $y_c(=2a_+)$,
gives a non-trivial scaling limit in the case of finite $y$. Furthermore
for general $k'<k$ we derive the equation satisfied by $u_1$ and argue that
the equation \eqref{yinfscaledstr2} is still valid. Each of these results
are also derived in Appendix \ref{HPrel} in a more rigorous manner.

For a particular potential $V_{k,k'}$ we could of course follow \cite{SN} and
insert \eqref{scaleansatz} and
%, \eqref{rsscaling1} and
\eqref{rsscaling2} into either the flow or string equation \eqref{floweqn}
or \eqref{stringeqn} and then Taylor expand in order to derive the relevant
differential equations for the scaling functions $u_i$ and $v_i$.  However
for general $k$ we require a general method that can uncover the structure
of such a hierarchy of equations. The particular approach we adopt here is
to follow the method of Appendix \ref{standarddsl} and solve \eqref{stringeqn}
using pseudo-differential operators. For a review of this technique we refer
to \cite{PDFGZ} as well as to \cite{Das} for its relation to the Lax method.

%%%%%%%%%%%%%%%%%%%%%%%%%%%%%%%%%%%%%%%%%%%%%%%%%%%%%%%%%%%%%%%%%%%%%%%%%%%%
%%%%%%%%%%%%
\subsection{Scaled String Equation and Lenard Differential Operators}
\label{stringL}

In the large-$N$ d.s.l our matrices $B,\ P$ and $H$ which all have only a
finite number
non-zero off diagonals will scale to differential operators of finite order.
The commutation relations that we derived in the previous section will then
allow us to directly obtain a differential equation for the scaling function.
First consider the scaling limit of the matrix $B$. Using our general scaling
ansatz \eqref{scaleansatz} it scales for all $k$ to \cite{PDFGZ},
\beq
B_{nm}\pi_m &\rightarrow& \left(B_c + a \sqrt{r_c} \left(d^2 + u_1(x) \right)
+ O(a^2)\right)\pi_n
%= B_c + a \sqrt{_c} + O(a^2),
\label{Blim}
\eeq
where $B_c=\sqrt{r_c}(2+s_c)$ is a constant operator and
we have defined $d  \equiv d/dx$. For later convenience we also define
\beq
\C{B} \equiv d^2 + u_1(x)\ .
\label{CBdef}
\eeq
The fact that these two terms contributing to $\C{B}$ are of the same order
fixes the scaling relation in \eqref{scaleansatz} in $\xi$ and $N$ as
functions of $a$.

For a given potential the matrix $H$ will scale to a differential operator of
a fixed degree, say $m$. Assuming that $H$ has a definite scaling dimension
it will be fully determined by the degree $m$,
\beq
H \rightarrow a^{(m-2k-1)/2} c_1 c_2 \left(d^m + \ldots \right)
+ O(a^{(m-2k-1)/2+1}) \equiv a^{(m-2k-1)/2} \C{H} + O(a^{(m-2k-1)/2+1}),\
\eeq
where $c_2$ is a constant. Here we have used the result from the previous
calculation that every power of $d$ comes with a scaling dimension $a^{1/2}$,
and we have multiplied by the extra power of $N$ in front of the potential
in \eqref{Hfinal}. We may therefore write \eqref{stringeqn} $[B,H]=B-y$ as
follows,
\beq
\label{streqn2}
a (a^{(m-2k-1)/2} [\C{B},\C{H}]+  O(a^{(m-2k-1)/2+1}))
= a (d^2 + u_1 - s) +O(a^2),
\eeq
where we have dropped the trivial constant part $B_c-y_c$ that has to be
satisfied. Here we have fixed $c_1=\sqrt{r_c}$ for convenience.
We see that in order to get a non-trivial scaling limit we require $m = 2k+1$.
To go further let
\beq
\C{H} \equiv \bar{\C{H}} + \frac{1}{4}\{x,d\}\ ,
\eeq
where $\{,\}$ denotes the anti-commutator. A short calculation shows that
\beq
\label{streqn3}
[\C{B},\bar{\C{H}}] = \frac{1}{2} x u_1'(x) + u_1(x) - s\ .
\eeq
The right hand side of \eqref{streqn3} is now a multiplicative operator and
we can therefore directly apply the technique of pseudo-differential operators
reviewed in \cite{PDFGZ} to find an expression for $\bar{\C{H}}$ such that the
left hand side of \eqref{streqn3} is also a multiplicative operator. Indeed
it is known that in general,
\beq
\label{genH}
\bar{\C{H}} = \sum^{k+1}_{n=1} b_n (\C{B}^{(2n-1)/2})_+,
\eeq
where here the subscript $+$ denotes the local part of the operator
$\C{B}^{n-1/2}$, which contains up to order $2n-1$ differential operators,
$d^{2n-1}$. The terms in \eqref{genH} for which $n < k+1$ will have a scaling
dimension incompatible with $\bCH$ unless the coefficients $b_n$ also come
with a scaling dimension. The only parameter with a scaling dimension in the
expression \eqref{Hexp} is $y$ and therefore the most general solution for
$\bar{\C{H}}$ is,
\beq
\label{specH}
\bar{\C{H}} = c_1 c_2 \left((\C{B}^{(2k+1)/2})_+
- b_k c_2^{-1} s (\C{B}^{(2k-1)/2})_+\right).
\eeq
This form which we have deduced here on the ground of general scaling
arguments is derived in Appendix \ref{HPrel} based on the relation
\eqref{Hexp} between $H$ and $P$ from the previous section. There it
is also shown that the relative coefficient, $b_k c_2^{-1} $, is unity.
Finally we can use the known fact that \cite{PDFGZ}
\beq
\label{BcomLenard}
[\C{B}, (\C{B}^{(2l-1)/2})_+] = -4^{1-l} \C{L}'_l[u_1]\ ,
\eeq
where $\C{L}_l$ is the Lenard differential operator and $\C{L}'_l$ denotes
its $d$-derivative.
The Lenard differential operator is related to the Gelfand-Dikii polynomials
$R_l$ by a trivial rescaling, $\C{L}_l=4^lR_l$. It is defined by the
following recursion relation
\footnote{Note that in the notation of \cite{PDFGZ} our $u_1(x)\to-u(x)$,
with $\C{B}=d^2+u_1\to Q=d^2-u$.},
\beq
\label{Lenardrelation}
\C{L}_{l+1}' [f] = d \C{L}_{l+1} [f] = \left(d^3 + 4 f d + 2f'\right)
\C{L}_l [f] \qquad \mathrm{and} \qquad \C{L}_0 = \frac{1}{2}\ .
\eeq
Because this only determines $\C{L}'_l$ one also requires the condition
that $\C{L}_l [0] = 0$ for all $l$. We may thus write \eqref{streqn3} as
\beq
-4^{-k} c_1 c_2\left( \C{L}_{k+1}'[u_1] - 4 s \C{L}_{k}'[u_1] \right) =
\frac{1}{2} x u_1'(x) + u_1(x) - s.
\eeq
 We now make use of the fact that under a rescaling of the variables,
$x \rightarrow \rho^{-1} x$, $u_1 \rightarrow \rho^{2} u_1$ and
$s \rightarrow \rho^{2} s$, we have $\C{L}_{l}[u_1]
\rightarrow \rho^{2l} \C{L}_{l}[u_1]$. Setting \footnote{Note here the minus sign is necessary to cancel the minus sign on the RHS of the preceeding equation. This extra minus will later appear in the final expression for the distribution by changing the integration region from $[0,\infty)$ to  $(-\infty,0]$},
\beq
\label{rescalefactor}
\rho = - 2^\frac{2k-1}{2k+1} (c_1 c_2)^\frac{-1}{2k+1},
\eeq
we obtain the equation,
\beq
\label{finalstreqn}
\C{L}_{k+1}'[u_1] - 4 s \C{L}_{k}'[u_1] = x u_1'(x) + 2u_1(x) - 2 s.
\eeq
This is the differential equation our scaling function has to satisfy for
the $k$th multicritical point. From \cite{Clarkson} we see it to be a
modification of the Painlev\'e XXXIV hierarchy. For the first few values of
$k$ it takes the form,
\bea
k = 1: &&  u_1^{(3)}+6 u_1 u_1'-4 s u_1' = x u_1'(x) + 2u_1(x) - 2 s, \\
k = 2: &&   u_1^{(5)}+10 \left(u_1' \left(3 u_1^2+2 u_1''\right)
+u_1 u_1^{(3)}\right)
%\nn\\
%&&  \qquad \qquad
-4 s \left(u_1^{(3)}+6 u_1 u_1'\right)
= x u_1'(x) + 2u_1(x) - 2 s,\nn\\
&&\\
k = 3: &&
 u_1^{(7)}+70 u_1''u_1^{(3)}+42u_1' u_1^{(4)}+14u_1 u_1^{(5)}+
70\left((u_1')^3+4u_1u_1'u_1''+u_1^2u_1^{(3)}\right)
+140u_1^3u_1'
\nn\\
&&
-4s\left(u_1^{(5)}+10 \left(u_1' \left(3 u_1^2+2 u_1''\right)
+u_1 u_1^{(3)}\right) \right)
= x u_1'(x) + 2u_1(x) - 2 s,
\eea
where we emphasise that $s$ plays the role of a parameter and all
differentials are with respect to $x$. Furthermore, from \eqref{Hfinal}
we may read off the d.s.l of $P$. We may write \eqref{Hfinal} as,
\beq
\label{splitH}
H=-\frac{N\al}{2}\Big( (V'(B)B)_+- (V'(B)B)_-\Big) - y P \ .
\eeq
so $y$ only appears in the term $-yP$. If we assume $P$ scales as
$P \rightarrow a^{-\Delta} \C{P}$, where $\Delta$ is its scaling
dimension we see that,
\beq
-yP \rightarrow -\left(y_c +c_1 s a\right)a^{-\Delta} \C{P}.
\eeq
In order for the term containing $s$ to match the term appearing in
\eqref{specH} we conclude $\Delta = 1$ and
$\C{P} = c_2 (\C{B}^{(2k-1)/2})_+ $. Note that the leading order term which
scales as $a^{-\Delta}$ must cancel against the other terms in \eqref{splitH}.
If we now substitute the d.s.l of $P$ together with the usual scaling ansatz
into the flow equation \eqref{floweqn} and use \eqref{BcomLenard} we obtain,
\beq
\label{scaledflow}
\partial_s u_1(x,s) = \partial_x\left(x - 2 \C{L}_k[u_1] \right).
\eeq
This equation will be useful in the next subsection
in order to prove equivalence
with the results of \cite{TC1}. This equation together with
\eqref{finalstreqn} are the main results of this subsection.

%%%%%%%%%%%%%%%%%%%%%%%%%%%%%%%%%%%%%%%%%%%%%%%%%%%%%%%%%%%%%%%%%%%%%%%%%%%%%%
\subsection{Higher Order Analogues of Tracy Widom and Examples}

Now we consider the d.s.l of the expression \eqref{logZ}. First
we take $N \to \infty$ to obtain,
\beq
\log[ Z_N(y; \alpha, \{g_l\})] = N^2 \int^1_0 d\xi
\left (1-\xi \right)
\log[r(\xi,y)] + O(1/N)\ .
\eeq
Note that this formula justifies setting $\xi_c = 1$; any non-analyticity
of $r(\xi,y)$ will only affect the partition function if it occurs in the
range $[0,1]$, hence as we approach the critical point we must zoom in on
the end point of the integration region \cite{PDFGZ}.  If we now substitute
the scaling ansatz for general $k$, the above takes the form,
\beq
\log Z^{(k)}(s) = -\int^0_{-\infty} dx \frac{x}{2} \left(u_1(x,s) + v_1(x,s)
\right)\ ,
\eeq
where we have introduced the scaled partition function $\log Z^{(k)}(s)$ for the
$k$th multicritical point and the limits have acquired an extra minus sign
due to the rescaling \eqref{rescalefactor}. In \eqref{finalstreqn} we have
only an equation for $u_1(x,s)$ and so must specify $v_1(x,s)$. In the Appendix we show that the $v_1$ in fact satisfies \eqref{yinfscaledstr2} for finite $y$. This behaviour is physically resonable as this equation
depends only on the behaviour of the eigenvalue density at the end of the
support at which the infinite wall is not imposed. Since the modification due to $y$ being
finite only affects the local behaviour of the eigenvalue density at
the other end of the support this will not have an effect at the opposite
end and therefore we expect \eqref{yinfscaledstr2} still holds. Imposing
the same boundary conditions on $v_1$ as the $y \to \infty$ case, we have
for the double scaling limit of the gap probability,
\beq
\label{gapProb}
\log \BB{P}^{(k)}(s) = -\int^0_{-\infty} dx \frac{x}{2}
\left(u_1(x,s) - u_1(x,\infty) \right)\ ,
\eeq
where $u_1(x,\infty)$ is the special function appearing in the expression
\eqref{yinfscaledstr1} i.e. it solves the $k$th member of the Painlev\'e I
hierarchy. The expression \eqref{gapProb} together with \eqref{finalstreqn}
constitute the main result of this paper; they give an expression for the
$k$th multicritical
gap probability and hence higher order analogues of the TW
distribution. Indeed we
can give a useful expression for the distribution of the largest eigenvalue
using \eqref{gapProb}, we have,
\beq
\label{dgapProb}
\frac{d}{ds} \log \BB{P}^{(k)}(s) = - \int^0_{-\infty}  dx \frac{x}{2}
\partial_s u_1(x,s).
\eeq
Using \eqref{scaledflow} in \eqref{dgapProb} we have,
\beq
\frac{d}{ds} \log \BB{P}^{(k)}(s) = -\int^0_{-\infty}dx \left(\C{L}_k[u_1]
- \half x \right).
\eeq
where we have integrated by parts assuming that the boundary conditions
on $u_1$ cause the boundary terms to vanish. The requirement that the
boundary terms vanish implies that as $x \to -\infty$ we have
$u_1(x,s) \rightarrow u_1(x,\infty) + \delta u_1(x,s)$, where
$\delta u_1 (x,s)$ is a subleading contribution containing the
$s$ dependence. We note here that this boundary condition is
compatible with \eqref{finalstreqn} since $u_1(x,\infty)$ is in
fact a solution to \eqref{finalstreqn}. We do not discuss the
boundary condition for $\delta u_1(x,s)$ here, instead we appeal
to \cite{TC1} in which a proposal was made for the appropriate
asymptotic behaviour of the solution. In the next subsection we
relate our results to theirs; one could then in principle convert
their boundary conditions into a condition on $\delta u_1(x,s)$.

As an example we now reproduce the standard Tracy-Widom law. For $k=1$ we
have that,
\beq
\C{L}_1[u_1(x,\infty)] = \frac{x}{2}\ \Rightarrow\ u_1(x,\infty) = \frac{x}{2}
\ ,
\eeq
and $u_1(x,s)$ satisfies \eqref{finalstreqn} for $k = 1$ which can be
written as,
\beq
\label{TWueqn}
u_1^{(3)}+6 u_1 u_1'-(x+4 s) u_1' - 2u_1 + 2 s = 0\ ,
\eeq
where all differentials are with respect to $x$. Although the choice of
normalisation for $u_1$, $x$ and $s$ in the main text gives the cleanest
results for general $k$, it differs from the choice made by Tracy and Widom
\cite{TW}.
We therefore perform some rescaling, in particular let
$u_1(x,s) = \gamma^2 \bar{u}(\gamma x, \gamma^{-2} s)$ with
$\gamma = -2^{-1/3}$. Then \eqref{TWueqn} becomes,
\beq
\label{TWueqn2}
\bar{u}^{(3)}(x,s)+2 \bar{u}(x,s)( 2 + 3 \bar{u}'(x,s))+2(x-2 s)
\bar{u}'(x,s) - 4s = 0\ .
\eeq
Finally, following \cite{SN}
let $-2q(x,s)^2 = \bar{u}(x,s) + x$, substituting this into
the above we find firstly that $q(x,s) = q(x+s,0) \equiv q(x+s) $
and secondly it may be written in the form,
\beq
q(x) W'(x) = -3 q'(x) W(x)\ ,
\eeq
where $W(x) \equiv q''(x) - 2q(x)^3 - x q(x)$. This then can be
integrated to show that $q(x)$ satisfies Painlev\'e II with $\alpha=0$.
We now may
write \eqref{gapProb} as,
\bea
\log \BB{P}^{(k)}(s) &=& -\int^0_{-\infty} dx \frac{x}{2} \left(u_1(x,s)
- \frac{x}{2}\right)
%, \nn \\
=
-\int^0_{-\infty} dx \frac{x}{2} \left(\gamma^2 \bar{u}(\gamma x,
\gamma^{-2}s) - \frac{x}{2}\right).\ \
\eea
Making the change of variables $\bar{x} = \gamma x$ we find,
\bea
\log \BB{P}^{(k)}(s) &=& \int^{\infty}_0 d\bar{x}\ \frac{\bar{x}}{2}
\left(\bar{u}(\bar{x},\gamma^{-2}s) + \bar{x}\right)
%, \nn \\
%&=&
=-\int^{\infty}_0 dx\ x q(x+\gamma^{-2}s)^2, \nn \\
\Rightarrow\  \log \BB{P}^{(k)}(\gamma^{2}s)
&=& \int^{\infty}_{s} dx (s-x) q(x)^2\ ,
\eea
which coincides with the TW distribution.

%%%%%%%%%%%%%%%%%%%%%%%%%%%%%%%%%%%%%%%%%%%%%%%%%%%%%%%%%%%
\subsection{Relation to Painlev\'e II via B\"{a}cklund Transformations}

In this section we relate our string equation \eqref{finalstreqn} to the
Painlev\'e II hierarchy, thereby making contact with the results of \cite{TC1}.
We proceed using the B\"{a}cklund transformations introduced in \cite{Clarkson}
which we now review.

We consider two B\"{a}cklund transformations defined by the equations,
\beq
\label{BT1}
2 \C{L}_k[u_1] - x = 2 \psi(x)^2 \qquad \mathrm{and}
\qquad \psi''(x) + (u_1- s) \psi(x) = 0\
\eeq
and
\beq
\label{BT2}
2 \C{L}_k[W' - W^2 + s] - x = 2 \psi(x)^2 \qquad \mathrm{and}
\qquad \psi'(x) + W \psi(x) = 0 .
\eeq
In the above equations we have introduced an auxilary variable $\psi$ which will
prove convient for comparison with \cite{TC1}. However, ultimately we are interested
in the composition of the above B\"{a}cklund transformations in order to obtain a relation
between the quantity $u_1$ and the new variable $W$.

We now consider the relation between $u_1$ and $\psi$ generated by \eqref{BT1},
by eliminating $u_1$ and $\psi$ respectively, as done in \cite{Clarkson}. Eliminating $\psi$ in \eqref{BT1} yields,
\beq
K_n K_n''-\frac{1}{2} (K_n')^2 + 2(u_1 -s)K_n^2 = 0\ ,
\eeq
where we have introduced the useful quantity $K_n[u_1] = 2 \C{L}_k[u_1] -x$.
Differentiating this we obtain,
\beq
\label{finalstreqnBT}
(d^3 + 4u_1 d + 2u_1) \C{L}_k[u_1] - 4s \C{L}'_k[u_1] = x u_1' + 2u_1 -2s\ ,
\eeq
which using the Lenard recursion formula gives \eqref{finalstreqn}.
On the other hand eliminating $u_1$ gives,
\beq
\label{psieqn}
2\C{L}_k\left[s-\frac{\psi''}{\psi}\right] - x = 2 \psi^2(x)\ .
\eeq
The solutions of \eqref{finalstreqnBT} and \eqref{psieqn} are therefore related by the B\"{a}cklund transformation \eqref{BT1}.

We now perform the same calculation using \eqref{BT2}. Eliminating $W$ yields again \eqref{psieqn}, whereas eliminating $\psi$
gives,
\beq
\label{Weqn}
(d + 2W) \C{L}_k[W' - W^2 + s] = xW+\frac{1}{2}\ .
\eeq
The solutions of \eqref{psieqn} and \eqref{Weqn}, and hence the solutions of \eqref{finalstreqnBT} and \eqref{Weqn}, are therefore related by the B\"{a}cklund transformation \eqref{BT1} and \eqref{BT2}. The equation \eqref{Weqn} is closely related to the Painlev\'e II equation with $\alpha = -1/2$. We can make this connection more explicit by using
the lemma,
\beq
\C{L}_k[u(x) + z] = \sum^k_{j = 0} (4z)^{k-j}
\frac{\Gamma(k+1/2)}{\Gamma(k-j+1)\Gamma(j+1/2)} \C{L}_j[u(x)]\ ,
\eeq
where $z$ is a constant. This lemma can be verified by induction
as it is shown in Appendix \ref{LA} using the
Lenard recursion relation and seems not to have appeared in the literature
before. Hence \eqref{Weqn} can be written,
\beq
(d + 2W) \C{L}_k[W' - W^2 ] + \sum^{k-1}_{j = 0} \tau_j\left(
2^{(4k-2)/(2k+1)} s\right)
(d + 2W) \C{L}_j[W' - W^2 ] = xW+\frac{1}{2}\ ,
\eeq
where
\beq
\tau_j(s) = (2j+1) 2^{(2k-4j-1)/(2k+1)}
 \frac{\Gamma(k+1/2)}{\Gamma(k-j+1)\Gamma(j+3/2)} s^{k-j}
\eeq
corresponds to the $\tau_j$ in \cite{TC1} if $k$ is odd, and the set of
parameters $t_l$ there is set to zero.
In order to study the distribution of the
largest eigenvalue the insertion of such parameters is not necessary, and we
refer to \cite{PDFGZ} for the interpretation of these terms.

Let us introduce the function,
\beq
U(x,s) = -\psi(x + \tau_0(\beta s),s)\ ,
\eeq
where $\beta = 2^{(4k-2)/(2k+1)}$. Using the B\"{a}cklund transformations
we may write the largest eigenvalue distribution as,
\bea
\frac{d}{ds} \log \BB{P}^{(k)}(s) &=&
-\int^0_{-\infty} dx\left(\C{L}_k[u_1] - \half x \right)
= \int^0_{-\infty}dx\ U(x-\tau_0(\beta s),s)^2 \nn\\
&=& \int^{\, -\tau_0(\beta s)}_{-\infty}dx\ U(x,s)^2 \ ,
\eea
and we have that $U'(x,s)/U(x,s) = W(x + \tau_0(\beta s), s)$ with
$W(x,s)$ satisfying \eqref{Weqn}.
Letting $q(x,s) = -W(x + \tau_0(\beta s),s)$ as well as
using the known symmetry
of the Painlev\'e equations (see e.g. \cite{Clarkson})
we obtain the equations appearing in Theorem 1.12 of \cite{TC1} in $q$ with $\alpha=+\frac12$
%we have that $q(x,s)$ satisfies,
%\beq
%(d + 2q) \C{L}_k[q' - q^2 ] + \sum^{k-1}_{j = 1} \tau_k(\beta s)
%(d + 2q) \C{L}_k[q' - q^2 ] = xq-\frac{1}{2}, \nn
%\eeq
and we have the relation,
\beq
U''(x,s)/U(x,s) = q'(x,s) + q^2(x,s)\ .
\eeq
This exactly reproduces the results of \cite{TC1}. Note that we make the identification of our $\tau_0(s)$ with the function 
$-x(s)$ appearing in \cite{TC1}.

%%%%%%%%%%%%%%%%%%%%%%%%%%%%%%%%%%%%%%%%%%%%%%%%%%%%%%%%%%%%%%%%%%%%%%%%%%%
\sect{Conclusions and Outlook}\label{conc}

In this paper we have presented an alternate derivation of the higher
order analogues of the Tracy-Widom distribution
which are characterised by the spectral density vanishing as a
$(2k-1)/2$-root.
Instead of using Fredholm determinants
we apply orthogonal polynomials to directly
calculate a truncated partition function for the matrix model in which the
upper limit of the integration over the eigenvalues is finite. The gap
probability for the matrix model can then be expressed in terms of this
truncated partition function. This work therefore directly extends an
earlier paper \cite{SN} in which the usual Tracy-Widom distribution was
derived. Our results should also be compared with those obtained earlier
in \cite{TC1} via Riemann-Hilbert methods. We have shown the results
presented here are equivalent to those appearing in \cite{TC1}
for odd values of $k$ via a
sequence of B\"{a}cklund transformations. One pleasing aspect of our
approach is that the resulting higher order analogues of Tracy-Widom
can be stated more succinctly than in \cite{TC1}. Indeed we have,
\beq
\log \BB{P}^{(k)}(s) = -\int^0_{-\infty} dx \frac{x}{2}
\left(u_1(x,s) - u_1(x,\infty) \right)
\eeq
where $\BB{P}^{(k)}(s)$ is the gap probability at the $k$th multicritical
point for $k\in\mathbb N$, and $u_1(x,s)$ satisfies,
\beq
\C{L}_{k+1}'[u_1] - 4 s \C{L}_{k}'[u_1] = x u_1' + 2u_1 - 2 s\ ,
\eeq
where $\C{L}_l$ is the Lenard differential operator, see e.g. \cite{PDFGZ}.
Here all derivatives denoted by prime are with respect to $x$. The issue of what boundary
conditions to impose on $u_1$ are discussed in the main text but are only
given implicitly in terms of the boundary conditions chosen in \cite{TC1}.

%It should be noted that the result presented in \cite{TC1} is more general
%in that they scale the parameters $\{t_l\}$ appearing in the potential so
%that they also appear in the d.s.l. Our result should be compared to theirs
%when all such parameters are zero. The obvious generalisation of our result
%in the case of non-zero $t_l$ would be for $u_1(x,s,{t_l})$ to satisfy the
%equation,
%\beq
%\C{L}_{k+1}'[u_1] - 4 s \C{L}_{k}'[u_1] + \sum^{k}_{l=1} t_l \C{L}_{l}'[u_1]
%= x u_1' + 2u_1 - 2 s.
%\eeq
%We give the equation above as a conjecture; one would need to generalise
%the B\"{a}cklund transformations appearing here and in \cite{Clarkson}
%further in order to make contact with \cite{TC1}. In particular the mapping
%between the $t_l$ parameters in each paper is not obvious.

Let us emphasize that for technical reasons we had to restrict ourselves to a spectral density with one-arc support. However, we conjecture that the same generalised Tracy-Widom distributions will appear at suitably tuned inner or outer edges of a multi-arc support.

%It is worth noting that we obtain the gap probability rather than the
%distribution of the largest eigenvalue as done in \cite{TC1}. One suspects
%then that the issue of the integration constant raised in \cite{TC1} could
%be addressed more easily in our approach.

Finally, one advantage of avoiding the use of Fredholm determinants is that
one should be able to calculate the probability distribution for
large-deviations of the maximum eigenvalue from its mean. Such large
deviations were the main focus of \cite{SN} and since our method is a
direct extension of theirs we expect that such an analyse could be
performed for the higher order cases. With our approach one could also
investigate issues of universality in the large deviation tails. This
would also give an alternative to the method in \cite{BN} which
appears to also be able to address these questions. This is something
we hope to pursue in future work.

%Finally it would be interesting to find some physical applications of the distributions presented here. Given that the multicritical limit of matrix models are known to be associated to $(p,q)$ minimal models

{\bf{Note}}: While completing this work an interesting paper appeared
\cite{Adler2} in which partial differential equations for Fredholm
determinants associated to the d.s.l of one and two matrix models were
constructed using string equations and the method of pseudo-differential
operators. Since the gap probability may be expressed as a Fredholm
determinant there is a large overlap in the results presented here with
those appearing in \cite{Adler2}. However there are some important
differences; firstly we give an explicit expression for the gap probability
in terms of a solution to Painlev\'e XXXIV; hence our expression is in some
sense a solution to the PDEs derived in \cite{Adler2}. However, on this
point, it is not immediately clear how the PDEs appearing in \cite{Adler2}
compare to our expressions as they seem to require the introduction of
extra coupling constants in the potential which also scale non-trivially
in the d.s.l. Finally, the B\"{a}cklund transformation appearing here
seems to partially answer the question raised in \cite{Adler2}
concerning the relation between their approach and that of \cite{TC1}.\\

%\ \\[1ex]

{\bf Acknowledgments:}
We acknowledge partial support
by the SFB $|$ TR12 ``Symmetries and Universality
in Mesoscopic Systems'' of the German research council DFG (G.A.).

\begin{appendix}
\sect{Equivalence of Gaussian Flow Equations
%for the  Gaussian Potential
}\label{stringequiv}

The aim of this section is to prove the equivalence between our set of
flow equations \eqref{snG} and \eqref{rnG} for the Gaussian potential
%that relate the recurrence relations
%$r_n,\ s_n$
and
a respective set of recurrence relations that was derived in ref. \cite{SN}
%For convenience we repeat here our results
%\bea
%s_{n} - s_{n-1} &=& - \frac{1}{2\alpha N} \partial_y \log r_{n}
%\label{snGapp}\\
%r_{n+1} - r_{n} &=& \frac{1}{2\alpha N}(1-\partial_y s_n)\ ,
%\label{rnGapp}
%\eea
in eqs. (42) and (41) there:
\bea
s_{n} - s_{n-1} &=& - \frac{1}{2\hal} \partial_y \log [r_{n}]\ ,
\label{snGSN}\\
r_{n+1} - r_{n-1} +s_n^2-s_{n-1}^2&=& -\partial_\hal \log[r_n]\ ,
\label{rnGSN}
\eea
with $\hal=N\al$.
Obviously the first equation agrees with \eqref{snG}.
To show that eq. \eqref{rnGSN} also follows from our formalism consider first
the matrix $P$ in the form of eq. \eqref{Pdef}. Due to its antisymmetry the
diagonal has to vanish,
\beq
P_{nn}&=&A_{nn}+C_{nn}-\hal B_{nn}=-\frac12\partial_y\log[h_n]-\hal s_n=0\nn\\
\Rightarrow\ \ s_n&=&-\frac{1}{2\hal}\e^{-\hal y^2}\pi_n(y)^2\ ,
\label{sn}
\eeq
where we have used \eqref{Alimits} and \eqref{Cnn} or equivalently
\eqref{id1} in the second line.
Our main relation is obtained from the lower diagonal,
\beq
P_{n,n-1}&=&A_{n,n-1}+C_{n,n-1}-\hal B_{n,n-1}=
\frac{n}{\sqrt{r_n}}-\e^{-\hal y^2}\pi_n(y)\pi_{n-1}(y)-\hal\sqrt{r_n}\nn\\
&=&
-P_{n-1,n}=\hal\sqrt{r_n}\nn\\
\Leftrightarrow n&=&\sqrt{r_n}\e^{-\hal y^2}\pi_n(y)\pi_{n-1}(y)+2\hal r_n\ ,
\label{Pid}
\eeq
using eqs. \eqref{Aid} and \eqref{id1} as well as the antisymmetry
of $P$. We can
now use the three step recurrence relation
to eliminate $\sqrt{r_n}\pi_{n-1}(y)$ as
well as use \eqref{sn}:
\beq
n&=&\e^{-\hal y^2}\pi_n(y)\left((y-s_n)\pi_n(y)-\sqrt{r_{n+1}}\pi_{n+1}(y)
\right)+2\hal r_n\nn\\
&=&-2\hal(y-s_n)s_n-(n+1-2\hal r_{n+1}) +2\hal r_n\ ,\nn\\
\Leftrightarrow 2n+1&=&2\hal((s_n-y)s_n+ r_{n+1}+ r_n )\ ,
\label{Pid2}
\eeq
where we have also applied \eqref{Pid} for $n\to n+1$. Taking the difference
between eq. \eqref{Pid2} at $n$ and $n-1$ we finally arrive at
\beq
2&=&2\hal(y(s_{n-1}-s_n)+s_n^2-s_{n-1}^2+r_{n+1}-r_{n-1})\nn\\
&=&2\hal(y\frac{1}{2\hal} \partial_y \log [r_{n}]+s_n^2
-s_{n-1}^2+r_{n+1}-r_{n-1})
\ ,
\eeq
upon inserting eq. \eqref{snGSN}. The last step to arrive at  eq. \eqref{rnGSN}
requires the following identity:
\beq
1-\frac{y}{2}\partial_y \log r_{n}=-\hal \partial_\hal \log[r_n]\ .
\label{yal}
\eeq
It follows by comparing the $y$- and $\hal$- derivative of
\beq
1=\int^y_{-\infty} d\lambda \e^{-\hal\lambda^2} \pi_{n}(\la)^2
=\frac{1}{\sqrt{\hal}}\int^{\sqrt{\hal}y}_{-\infty} dz \e^{-z^2}
\left(\frac{1}{\sqrt{h_n}}\left(\frac{z}{\sqrt{\hal}}\right)^n
+ O(z^{n-1})
\right)^2\ .
\eeq
For the former we have already from \eqref{sn}
\beq
\partial_y \log[h_n]=\e^{-\hal y^2}\pi_n(y)^2\ ,
\eeq
whereas for the latter we obtain
\beq
0=-\frac{1}{2\hal}+\frac{y}{2\hal}\e^{-\hal y^2}\pi_n(y)^2
-\partial_\hal\log[h_n]-2\frac{n}{2\hal}\ .
\eeq
Inserting these equations into each other and taking the difference
$\log[h_n]-\log[h_{n-1}]=\log[r_n]$ we arrive at \eqref{yal}.

%%%%%%%%%%%%%%%%%%%%%%%%%%%%%%%%%%%%%%%%%%%%%%%%%%%%%%%%%%%%%%%%%%%%%%%%
%%%%%%%%%%%%%%%%%%%%%%%%%%%%%%%%%%%%%%%%%%%%%%%%%%%%%%%%%%%%
\sect{Multiplying Upper and Lower Triangular Matrices}\label{tri}

Any matrix $D$ can be decomposed into its strictly upper (+), diagonal (d),
and strictly lower triangular ($-$) part:
\beq
D=D_++D_{\rm d}+D_-
\eeq
The aim of this appendix is to express the + and $-$ part of the product of
two matrices in terms of the  +, d and $-$ parts of the individual
factors, in order to express matrix $H$ in terms of matrix $P$ in the main
text eq. \eqref{Hexp}.
Obviously multiplying any matrix by the diagonal part of a matrix $D_{\rm d}$
will not change the
character of that matrix to be +, d, or $-$:
\beq
(BD_{\rm d})_+=B_+D_{\rm d}\ ,\ \ (BD_{\rm d})_{\rm d}=B_{\rm d}D_{\rm d}
\ ,\ \ (BD_{\rm d})_-=B_-D_{\rm d}\ .
\label{diag}
\eeq
Furthermore multiplying two strictly upper (lower) triangular matrices gives
a strictly upper (lower) triangular matrix, $(B_+D_+)_+=B_+D_+$ and
$(B_-D_-)_-=B_-D_-$.

>From now on we will use that our matrix $B$ eq. \eqref{Bnmdef} is tridiagonal.
Therefore it can only shift off-diagonals at most up or down by one, that is
\beq
%BD_+&=&
%%(BD_+)_++(BD_+)_{\rm d}+(BD_+)_-=
%B_+D_++B_{\rm d}D_++B_-D_+\nn\\
%\Rightarrow \
(BD_+)_+&=&B_+D_++B_{\rm d}D_++(B_-D_+)_+=
BD_+-(B_-D_+)_{\rm d}\ ,
%\ \ \nn\\
%(BD_+)_{\rm d}&=&(B_-D_+)_{\rm d}\ ,\ \ (BD_+)_-=0\ .
\eeq
due to $(BD_+)_{\rm d}=(B_-D_+)_{\rm d}$ and $(B_-D_+)_-=0$.
%In
%%the last equation in
%the first line the first two terms are +, whereas
%the last term has a $d$ and a + part, and there is no - part. This leads to
%the identities in the second line.
Analogously one can deduce
\beq
%BD_-&=&
%%(BD_-)_++(BD_-)_{\rm d}+(BD_-)_-=
%B_+D_-+B_{\rm d}D_-+B_-D_-\nn\\
%\Rightarrow \
(BD_-)_-&=& (B_+D_-)_-+B_{\rm d}D_-+B_-D_-=
BD_--(B_+D_-)_{\rm d}\ ,
%\nn\\
%(BD_-)_{\rm d}&=&(B_+D_-)_{\rm d}\ ,\ \ (BD_-)_+=0\ .
\eeq
because of $(BD_-)_{\rm d}=(B_+D_-)_{\rm d}$ and $(B_+D_-)_+=0$.
Consequently we can write down for any product $BD$ its respective +, d
and $-$ part:
\beq
(BD)_+&=&BD_+-(B_-D_+)_{\rm d}
%B_+D_++B_{\rm d}D_++(B_-D_+)_+
+B_+D_{\rm d}\ ,\nn\\
(BD)_{\rm d}\ &=& (B_+D_-)_{\rm d}+B_{\rm d}D_{\rm d}+(B_-D_+)_{\rm d}\ ,\nn\\
(BD)_-&=&BD_--(B_+D_-)_{\rm d}
%(B_+D_-)_-+B_{\rm d}D_-+B_-D_-
+B_-D_{\rm d}\ .
\eeq
We can now proceed expressing $H$ through $P$. Defining
\beq
F=-\frac{N\al}{2}V'(B)
\eeq
which commutes with $B$ and is symmetric as is $B$, we have
\beq
P=
%P_++P_-=
F_+-F_-\ ,
\eeq
which has no diagonal part, $P_{\rm d}=0$. Finally we obtain
\beq
H&=&(BF)_+-(BF)_--y(F_+-F_-)\nn\\
%&=& BF_+-B_-F_++B_+F_{\rm d}+(B_-F_+)_+
%-(BF_--B_+F_-+(B_+F_-)_-+B_-F_{\rm d}) -yP\nn\\
&=&(B-y)(F_+-F_-)-(B_-F_+)_{\rm d}+(B_+F_-)_{\rm d}+B_+F_{\rm d}
-B_-F_{\rm d}\nn\\
&=& (B-y)P-(BP)_{\rm d}+(B_+-B_-)F_{\rm d}\ ,
\eeq
where in the last step we have added trivially vanishing terms
$(B_-P_-)_{\rm d}=0=(B_+P_+)_{\rm d}$ and $(B_{\rm d}P)_{\rm d}=0$ in
order to express everything
in terms of $P$.

%%%%%%%%%%%%%%%%%%%%%%%%%%%%%%%%%%%%%%%%%%%%%%%%%%%%%%%%%%%%%%%
\sect{The Double Scaling Limit in the Case $y \to \infty$}\label{standarddsl}

Here we recall some standard results about the d.s.l of the
Hermitian one-matrix model whose partition function is
$Z_N(\infty; \alpha, \{g_l\})$. The equations \eqref{Znorm} and \eqref{Bdef}
hold in this case with now all quantities independent of $y$. The recursion
coefficients in the definition of \eqref{Bdef} are determined by the recursion
relations, also known as the string equations,
\bea
%\label{Zinfstr1}
\alpha V'(B)_{n,n+1} &=& n/N \ ,\nn\\
V'(B)_{nn} &=& 0\ . \label{Zinfstr2}
\eea
Note that these equations can be obtained from \eqref{Pdef} by setting
derivatives of $y$, and hence $C$, to zero and using \eqref{Aid}. By taking
the large $N$ limit we can rewrite these equations in terms of the continuous
variable $\xi = n/N$. It is known
(see e.g. \cite{Moore})
that for certain critical potentials in
which the eigenvalue density vanishes as a $(2k+1)/2$-root at the right
and a $(2k'+1)/2$-root at the left end of the support, respectively,
with $k'<k$, that substituting the ansatz \eqref{yinfscaleansatz}
%, \eqref{rsscaling1}
and \eqref{rsscaling2} into %\eqref{Zinfstr1} and
\eqref{Zinfstr2} results,
after the rescaling \eqref{rescalefactor}, in the
equations \eqref{yinfscaledstr1} and \eqref{yinfscaledstr2}.
It was then shown in \cite{DJM} that the correct solution to
\eqref{yinfscaledstr2} was  $v_1 = 0$. The equations
%\eqref{Zinfstr1}
\eqref{Zinfstr2} may also be written as,
\beq
\label{opstreqn}
[B,P] = 1\ ,
\eeq
where $P$ is defined by \eqref{Pfinal}. The equations \eqref{yinfscaledstr1}
and \eqref{yinfscaledstr2} arise in this operator formalism since it is easy
to show that $B$ scales to a differential operator independent of $k$,
\beq
B \rightarrow B_c + a c_1 \left(d^2 + u_1(x) \right) + O(a^2) = B_c + a c_1
\C{B} + O(a^2)\ .
\eeq
Furthermore, due to the finite number of nonzero off-diagonals in $P$ and
its antisymmetry, we expect $P$ to scale to an anti-hermitian differential
operator.  In general for $k' < k$, $P \rightarrow a^{-\eta} \C{P} + \ldots$
where $\eta > 1$. Requiring that we have a scaling
$\eta = 1$ for the leading term determines the functions $v_i$, in particular
it gives \eqref{yinfscaledstr2}.
%Once we reach
%$P \rightarrow a^{-1} \C{P} + \ldots$ it is no longer possible to choose
%one of the $v_i$ to cancel this order, instead we get an equation for $u_1$.
Given that $[\C{B},\C{P}]$ must be a multiplicative operator we have
$\C{P} \propto (\C{B}^{2k-1})_+$ and therefore \eqref{yinfscaledstr1},
up to rescalings.

%%%%%%%%%%%%%%%%%%%%%%%%%%%%%%%%%%%%%%%%%%%%%%%%%%%%%%%%%%%%
\sect{Derivation of the d.s.l for $H$}\label{HPrel}

In contrast to the main text in subsection \ref{stringL} we first derive here
the scaling of operator $P$, and then using the relation between $H$ and $P$
show the d.s.l for $H$ to be of the form \eqref{specH}.
We now consider the case of the potential in $Z_N(y; \alpha, \{g_l\})$ taking
the form of a critical potential of $Z_N(\infty; \alpha, \{g_l\})$. The left
hand side of the flow equation \eqref{floweqn} is exactly the left hand side
of \eqref{opstreqn} and the right hand side of \eqref{floweqn} has the same
scaling dimension as the right hand side of \eqref{opstreqn}. This means that
using the ansatz \eqref{scaleansatz}
%, \eqref{rsscaling1}
and \eqref{rsscaling2}, we have for $k' < k$ that $v_1$ satisfies
\eqref{yinfscaledstr2} and furthermore we can also conclude
\beq
P \rightarrow a^{-1} \C{P} + O(1) = a^{-1} c_2 (\C{B}^{(2k-1)/2})_+  + O(1)\ .
\eeq
The double scaling limit of the flow equation is then, using \eqref{BcomLenard}
\beq
\label{scaledflowunscaled}
1 - \partial_s u_1 = -4^{1-k} c_1 c_2 \C{L}'_{k}[u_1]\ ,
\eeq
which will be of use in the following. We now turn our attention to
$H$ in \eqref{Hdef} and prove,
\beq
\label{scaleH}
\C{H} = c_2 c_1 \left( (\C{B}^{(2k+1)/2})_+ -
s (\C{B}^{(2k-1)/2})_+ \right) + \frac{1}{4}\{x,d\}\ .
\eeq
Recall the expression \eqref{Hexp}, we have,
\beq
H= \frac{1}{2}\{B,P\} - y P - \frac{1}{2}\left\{B_+ - B_-,
\frac{N\alpha}{2}[V'(B)]_{\rm d} \right\} \ .
\label{HPrelate}
\eeq
If we now consider the d.s.l it is a simple matter to
prove that independent of $k$,
\beq
B_+ - B_- \rightarrow -2c_1 a^{1/2} d + O(a).
\eeq
To find the double scaling limit of the operator
$\frac{N\alpha}{2}[V'(B)]_{\rm d}$ we must work slightly harder.
Recalling the expression for $P$ \eqref{Pdef} and \eqref{Cnn}, then
the antisymmetry of $P$ implies,
\beq
C_{nn} = -\partial_y \log[\sqrt{h_n}] = \frac{N\alpha}{2} V'(B)_{nn}\ ,
\eeq
which gives,
\beq
-\frac{1}{2}\partial_y \log[r_n] = \frac{N\alpha}{2} V'(B)_{nn}
- \frac{N\alpha}{2} V'(B)_{n-1,n-1}\ .
\eeq
where $V'(B)_{nn}$ are trivially related to the operator $[V'(B)]_{\rm d}$.
In the large $N$ limit we have
$\frac{N\alpha}{2} V'(B)_{nn} \rightarrow \hat{V}(\xi,y)$, where
$\hat{V}(\xi,y)$ is a continuous function of $\xi$. We therefore have,
\beq
-\frac{1}{2}\partial_y \log[r(\xi,y)] = \frac{1}{N}
\partial_\xi \hat{V}(\xi,y)\ .
\eeq
Substituting in \eqref{scaleansatz}
%, \eqref{rsscaling1}
and \eqref{rsscaling2} while assuming an arbitrary scaling dimension for
$\hat{V}$ i.e. $\hat{V}(\xi,y) \rightarrow a^\eta \C{V}(x,s)$ gives,
\beq
\frac{1}{4 c_1}\partial_s \left(u_1(x,s) + v_1(x,s) \right) =
\frac{1}{4 c_1}\partial_s u_1(x,s) = a^{\eta + 1/2} \partial_x \C{V}(x,s),
\eeq
from which we see see that $\eta = -1/2$. Furthermore, using
\eqref{scaledflowunscaled} we have,
\bea
\C{V}(x,s) = \frac{1}{4c_1} (x + 4^{1-k}c_1 c_2 \C{L}_{k}[u_1] ),
\eea
where we have integrated and assumed the integration constant is zero.

Returning now to \eqref{HPrelate}, the form of the d.s.l
for all operators is now known. Substituting them in we obtain,
\beq
\C{H} = \frac{1}{2}c_1 c_2 \{\C{B},(\C{B}^{k-1/2})_+\}
-c_1c_2s (\C{B}^{(2k-1)/2})_+
+ \frac{1}{4}\{x,d\} + 4^{-k} c_1 c_2 \{ \C{L}_{k}, d\},
\eeq
which upon using the identity \cite{PDFGZ}
$(\C{B}^{k+1/2})_+ =
\frac{1}{2} \{\C{B},(\C{B}^{k-1/2})_+\} +  4^{-k} \{ \C{L}_{k}, d\}$
becomes \eqref{scaleH}.

%%%%%%%%%%%%%%%%%%%%%%%%%%%%%%%%%%%%%%%%%%%%%%%%%%%%%%%%%%%%
\sect{Shift Identity for Lenard Differential Operators}\label{LA}

We claim that when $z$ is a constant it holds for any integer $k\in\mathbb N$:
\beq
\C{L}_k[u(x) + z] = \sum^k_{j = 0} (4z)^{k-j}
\frac{\Gamma(k+1/2)}{\Gamma(k-j+1)\Gamma(j+1/2)} \C{L}_j[u(x)]\ .
\eeq
We now prove this via induction. Firstly, it is trivially true for $k=1$.
Now let us assume,
\beq
\C{L}_k[u(x) + z] = \sum^k_{j = 0} (4z)^{k-j} \alpha^{(k)}_j \C{L}_j[u(x)]\ ,
\label{IA}
\eeq
where we have defined
\beq
\alpha^{(k)}_j \equiv \frac{\Gamma(k+1/2)}{\Gamma(k-j+1)\Gamma(j+1/2)}\ ,
\ \ \mbox{for}\ \ j=0,1,\ldots,k\ ,
\eeq
and for any $k\in\mathbb N$.
It is simple to see that the following identity holds for these constants:
\beq
\alpha^{(k+1)}_j \equiv \alpha^{(k)}_j + \alpha^{(k)}_{j-1}\ ,
\ \ \mbox{for}\ \ j=1,\ldots,k\ \ .
\eeq
Let us now substitute the assumption \eqref{IA} at $k+1$
into Lenard's recursion relation \eqref{Lenardrelation}:
\bea
\C{L}_{k+1}'[u+z] &=& \sum^k_{j=0} \left[ \alpha^{(k)}_j
(4z)^{k-j} \C{L}_{j+1}'[u] + (4z)^{k+1-j} \alpha^{(k)}_j \C{L}_j'[u] \right]
\nn \\
&=& \sum^{k+1}_{j=1} \alpha^{(k+1)}_j (4z)^{k+1-j} \C{L}_{j}'[u], \nn \\
\Rightarrow \C{L}_{k+1}[u+z]  &=& \sum^{k+1}_{j=1}
\alpha^{(k+1)}_j (4z)^{k+1-j} \C{L}_{j}[u] + C_{k+1} \ .
\eea
Here we have introduced $C_{k+1}$ as an integration constant and used the fact
that $\C{L}_0$ is a constant as well as that $\alpha_{k+1}^{(k+1)}=
\alpha_k^{(k)}=1$.
%\bea
%\alpha^{(k+1)}_j &=&  \frac{\Gamma(k+1+1/2)}{(k+1/2)\Gamma(k+1-j+1)
%\Gamma(j+1/2)}\left[\frac{\Gamma(k+1-j+1)}{\Gamma(k-j+1)}
%+ \frac{\Gamma(j+1/2)}{\Gamma(j-1%+1/2)}  \right], \nn\\
%&=&  \frac{\Gamma(k+1+1/2)}{\Gamma(k+1-j+1)\Gamma(j+1/2)}.
%\eea
To complete the proof by induction we must fix $C_{k+1}$. This can be done
by using the fact that $\C{L}_j[0] = 0$ for all $j>0$, so that we have,
\beq
\C{L}_{k+1}[z] &=& C_{k+1}\ .
\eeq
Due to the homogenous scaling of $\C{L}_{k+1}[z]$ we know there exists only
one term in $\C{L}_{k+1}[z]$ which is non-zero when $u(x) = z = const$; it
is, $\beta_{k+1} u^{k+1}$, where $\beta_{k+1}$ is a known constant that
follows from the properties of the Lenard differential. We therefore have,
\beq
C_{k+1} = \beta_{k+1} z^{k+1}.
\eeq
Let us finally determine $\beta_k$. This can be done by noting that in the
Lenard recursion relation, the only term in $\C{L}_k$ which contributes to
the $u^{k+1}$ term in $\C{L}_{k+1}$, is $\beta_k u^k$. Hence we have from
the Lenard recursion relation,
\bea
d(\beta_{k+1} u^{k+1}) &=& 4u \beta_{k} u^{k-1} u' + 2 u' \beta_{k} u^{k}
= 4\frac{k+\frac12}{k+1} \beta_{k} d(u^{k+1}), \\
\Rightarrow\ \beta_{k+1} &=& 4^{k}
\frac{\Gamma(k+3/2)}{\Gamma(k+2)\Gamma(3/2)} \beta_1 \ ,
\eea
where $\beta_1 = 1$.
Substituting the known value of
$C_{k+1}$ into our previous relations using $\C{L}_0 = 1/2$
gives,
\beq
\C{L}_{k+1}[u+z]  &=& \sum^{k+1}_{j=0} \alpha^{(k+1)}_j (4z)^{k+1-j}
\C{L}_{j}[u]\ ,
\eeq
thereby completing the proof.

 \end{appendix}
%%%%%%%%%%%%%%%%%%%%%%%%%%%%%%%%%%%%%%%%%%%%%%%%%%%%%%%%%%%%%%%%%%%%%%%%%%%%%

%%%%%%%%%%%%%%%%%%%%%%%%%%%%%%%%%%%%%%%%%%%%%%%%%%%%%%%%%%%%%%%%%%%%%%%%%%%%%

\end{document}